\documentclass[a4paper,twocolumn,amsfonts,amssymb,amsmath,preprintnumbers]{revtex4}

\usepackage[latin1]{inputenc}
\usepackage[T1]{fontenc}
\usepackage[english]{babel}
\usepackage{ae}

\usepackage{esvect}

\usepackage{graphicx}
\usepackage{dcolumn}
\usepackage{color}
\usepackage{bm}

\usepackage[pdftex]{hyperref}
\hypersetup{
  pdftitle={Limit on a Right-Handed Admixture to the Weak b -> c Current from Semileptonic Decays}, 
  pdfauthor={Robert Feger}, 
  pdfsubject={}, 
  pdfcreator={pdfTeX},
  pdfproducer={Robert Feger}, 
  pdfkeywords={},
}
\input{babarsym}

\newcommand{\cR}{\ensuremath{{c^\prime_\text{R}}}\xspace}
\newcommand{\cRn}{\ensuremath{{c_\text{R}}}\xspace}
\newcommand{\cL}{\ensuremath{c_\text{L}}\xspace}

\newcommand{\asympm}[2]{\genfrac{}{}{0pt}{1}{#1}{#2}}

\def\eg {e.\,g.\xspace}
\def\percent {\,\%\xspace}

\newcommand{\gevsq}{\ensuremath{\mathrm{\,Ge\kern -0.1em V^2}}\xspace}
\newcommand{\gevcube}{\ensuremath{\mathrm{\,Ge\kern -0.1em V^3}}\xspace}
\newcommand{\gevtothe}[1]{\ensuremath{\mathrm{\,Ge\kern -0.1em V^{#1}}}\xspace}

\def\X       {\ensuremath{X}\xspace}

\def\ellm    {\ensuremath{\ell^{-}}\xspace}

\def\porantip    {\kern 0.18em\optbar{\kern -0.40em p}{}\xspace}
\def\D       {\ensuremath{D}\xspace}

\newcommand{\semilepXc} { \ensuremath{\Bbar \rightarrow \X_{c} \ellm \nub }}

\newcommand{\semilepD} { \ensuremath{\Bbar \rightarrow \D \ellm \nu }}
\newcommand{\semilepDstar} { \ensuremath{\Bbar \rightarrow \Dstar \ellm \nu }}

\newcommand{\Gammasl}    {\ensuremath{\Gamma_{\mathit{SL}}}\xspace}

\newcommand{\mxmom}[1]  { \ensuremath{\langle m_{\X}^{#1} \rangle} }
\newcommand{\mxmomc}[1]  { \ensuremath{\langle (m_{\X}^{2} - \langle m_{\X}^{2}\rangle )^{#1}\rangle } }

\newcommand{\elmom}[1]  { \ensuremath{\langle E_{\ell}^{#1}\rangle } }
\newcommand{\elmomc}[1]  { \ensuremath{\langle (E_{\ell}-\langle E_{\ell}\rangle )^{#1}\rangle } }

\newcommand{\plep}         { \ensuremath{p_{\ell}^{*}} }
\newcommand{\plmin}      { \ensuremath{p_{\ell,\mathrm{min}}^{*}} }
\newcommand{\plgeq}[1]   { \ensuremath{\plep \geq {#1} \gevc }}

\newcommand{\plcut}      { \ensuremath{p_{\ell,\mathrm{min}}^{*}} }

\newcommand{\El}      { \ensuremath{E_{\ell}^{*}} }

\newcommand{\brf}        { \ensuremath{\mathcal{B}}}

\def\cov        { \ensuremath{\mathcal{C}}}

\newcommand{\mb}         { \ensuremath{m_{\b}}\xspace }
\newcommand{\mc}         { \ensuremath{m_{\c}}\xspace }

\newcommand{\mupi}      { \ensuremath{\mu_{\pi}^{2}}\xspace }
\newcommand{\muG}      { \ensuremath{\mu_{G}^{2}}\xspace }
\newcommand{\rhoD}      { \ensuremath{\rho_{D}^{3}}\xspace }
\newcommand{\rhoLS}      { \ensuremath{\rho_{\mathit{LS}}^{3}}\xspace }

\newcommand{\covtheo}{ \ensuremath{\cov_{\mathrm{theo}}}}

\newcommand{\covexp}{ \ensuremath{\cov_{\mathrm{exp}}}}
\newcommand{\covtot}{ \ensuremath{\cov_{\mathrm{tot}}}}

\begin{document}

\preprint{SI-HEP-2009-19}

\title{\boldmath
     Limit on a Right-Handed Admixture to the Weak $b \to c$ Current from Semileptonic Decays
} 

\author{Robert Feger}
\affiliation{Theoretische Physik 1, Fachbereich Physik, Universität Siegen, Walter-Flex-Str. 3, D-57068 Siegen, Germany}
\author{Verena Klose}
\author{Heiko Lacker}
\author{Thomas Lück}
\affiliation{Institut für Physik, Humboldt Universität zu Berlin, Newtonstr. 15, D-12489 Berlin, Germany}
\author{Thomas Mannel}
\affiliation{Theoretische Physik 1, Fachbereich Physik, Universität Siegen, Walter-Flex-Str. 3,  D-57068 Siegen, Germany}

\date{\today}

\begin{abstract}

We determine an upper bound for a possible right-handed $b{\to}c$ quark current admixture in semileptonic $\bar{B} \to X_{c} \ell^{-} \bar{\nu}$ decays from a simulateous fit to moments of the lepton-enery and hadronic-mass distribution measured as a function of the lower limit on the lepton energy, using data measured by the \babar\ detector. The right-handed admixture is parametrized by a new parameter $c_R$ as coefficient of computed moments with right-handed quark current. For the standard model part we use the prediction of the  heavy-quark expansion (HQE) up to order $1/m_b^3$ and perturbative corrections and for the right-handed contribution only up to order $1/m_b^2$ and perturbative corrections. We find $\cR=0.05\asympm{+0.33}{-0.50}$ in agreement with the standard-model prediction of zero. Additionally, we
give a contraint on a possible right-handed admixture from exclusive decays, which is with a value of $\cR =0.01{\pm}0.03$ more restrictive than our value from the inclusive fit. The difference in \Vcb between the inclusive and exclusive
extraction is only slightly reduced when allowing for a right-handed admixture
in the range of $\cR=0.01 {\pm} 0.03$.
\end{abstract}


\maketitle

\enlargethispage{5mm}

\section{Introduction}
\label{sec:introduction}

Parity violation is implemented in the Standard Model (SM) by assigning different
weak quantum numbers to left- and right-handed quarks and leptons and it is fair to
say  that there is yet no deeper understanding of the symmetry breaking mechanism in
weak interactions with respect to parity transformations. On the experimental
side, parity violation is well established in the leptonic sector, e.\,g. through the measurements
of the Michel parameters in the decay of Muons.

However, in hadronic transitions parity violation is much harder to test due to the
uncertainties  present in the calculation of the hadronic matrix elements of the  quark
currents. In turn, this leaves sizable room for a possible non-left-handed admixture.
While this is mainly true in the case of light quark systems, the calculational methods
have significantly developed for heavy quarks using the fact that the heavy quark masses
are large compared to the binding energy of heavy hadrons.

In particular for inclusive semileptonic $b{\to} c$ transitions the theoretical methods are
in a mature state and are applied in the framework of the SM to extract \eg the CKM Matrix
element $|V_{cb}|$ with an unprecedented relative precision of less than 2\percent \cite{Buchmuller:2005zv}.

Clearly the precision of the data as well as of the theoretical methods may serve also to
perform  a test for non-standard couplings. In two recent papers \cite{DFM1,DFM2} the
necessary calculations have been performed to check for a non-left-handed coupling
in inclusive semileptonic $b{\to}c$ transitions. In the present paper we use these results
together with the \babar\ data to obtain information on a possible right-handed admixture
to the weak $b{\to}c$ current.

Recently the tension between the inclusive and exclusive determinations
of $|V_{ub}|$ and---to a lesser extend---also of $|V_{cb}|$ motivated speculations
to explain this by right-handed admixtures in the weak hadronic currents.
In \cite{Crivellin:2009sd} it is shown that a right-handed admixture can soften the
tension and that a right-handed admixture can be obtained within the MSSM.

In the next section we recapitulate the theoretical input and fix our notation. In section~\ref{sec:analysis}
we perform the analysis based on \babar\ data. In section~\ref{sec:results} we discuss our result and compare them in
section~\ref{sec:exclusive} to limits from exclusive decays. Finally, in section~\ref{sec:summary} we conclude.
\section{Theory Background and Notation}
\label{sec:theory}

It is well known that any new physics effect beyond the SM can be parametrized in terms of
higher-dimensional operators, which are singlets under the SM symmetry
$\mathrm{SU}(3)_\text{QCD}{\times}\mathrm{SU}(2)_\text{weak} \times \mathrm{U}(1)_\text{Y}$. Assuming that the
Higgs sector is minimal (i.e.\ if one considers only a single Higgs doublet) there is only one
operator at dimension five which is related to a Majorana  mass of the neutrino and hence only affects the
leptonic sector. At dimension six one finds a long list of possible operators \cite{BuchWyl} among
which we find also operators modifying the helicity structure that appears in the semileptonic
$b{\to}c$ decays.

After spontaneous symmetry breaking
$\mathrm{SU}(3)_\text{QCD} \times \mathrm{SU}(2)_\text{weak} \times \mathrm{U}(1)_\text{Y} \to \mathrm{SU}(3)_\text{QCD}  \times \mathrm{U}(1)_\text{em}$
and after running down to the scale of the $b$-quark mass one finds for the effective interaction
\cite{DFM1,DFM2}
\begin{equation}\label{Hamiltonian2}
     \mathcal{H}_\text{eff} = \frac{4 G_\text{F} V_{cb}}{\sqrt2} J_{\text{q},\mu} J_\text{l}^\mu,
\end{equation}
where $J_\text{l}^\mu = \bar{e}\, \gamma^\mu P_-\, \nu_e$ is the usual leptonic current
and $J_{\text{h},\mu}$ is the generalized hadronic  $b{\to}c$ current  which is given by
\begin{equation} \label{EnhancedGamma}
     \begin{aligned}[t]
     J_{h,\mu}  &=   c_\text{L} \ \bar{c} \gamma_\mu P_- b
                   + c_\text{R} \ \bar{c} \gamma_\mu P_+ b\\
            &\quad + g_\text{L} \ \bar{c}\,i\overleftrightarrow{D_\mu} P_- b
                   + g_\text{R} \ \bar{c}\,i\overleftrightarrow{D_\mu} P_+ b \\[1mm]
            &\quad + d_\text{L} \ i \partial^\mu ( \bar{c}\, i \sigma_{\mu \nu} P_- b)
                   + d_\text{R} \ i \partial^\mu ( \bar{c}\, i \sigma_{\mu \nu} P_+ b) \, ,
     \end{aligned}
\end{equation}
where $P_\pm$ denotes the projector on positive/negative chirality and $D_\mu$ is
the QCD covariant derivative. Note that the leading term contributing to the rate
will be the interference term with the SM ($\propto c_\text{L}\cRn$), which means that the leptonic current
remains as in the SM since we consider only final states with electrons and muons and thus
can neglect the lepton mass,

Furthermore, \cL contains the SM contribution and hence
$\cL = 1+\mathcal{O}(v^2 / \Lambda^2)$, where $v$ is the vacuum expectation value from spontaneous symmetry breaking and $\Lambda$ is the new-physics' scale. All other contributions can only appear
through a new-physics effect.  In particular, the effective field theory approach
reveals  that $\cRn = \mathcal{O}(v^2 / \Lambda^2)$, while all helicity changing contributions
are expected to be further suppressed by a small Yukawa coupling \cite{DFM1, D'Ambrosio:2002ex}.

In the following analysis we restrict ourselves to an investigation of the parameter \cRn.
As has been shown in \cite{DFM2} the lepton-energy moments and hadronic-mass moments are
not very sensitive to the parameters $g_\text{L}$, $g_\text{R}$, $d_\text{L}$ and
$d_\text{R}$. Because the moments depend on the squared matrix element the parameters
appear in pairs, of which the leading contributions are $\cL^2$ and $\cL\cRn$. For the
combined fit the parameter \cL can also be dropped as an overall factor being absorbed in
\Vcb. Thus the parameter used in the fit is $\cR=\cRn/\cL$.


\section{Analysis}
\label{sec:analysis}

\subsection{Fit Setup}
The combined fit for the extraction of the new parameter \cR is performed along the lines as described in  \cite{Aubert:2007BABARHadronicMoments} using the HQEFitter package \cite{KloseSundermann}. It is based on the $\chi^2$ minimization,
\begin{equation}\label{eq:chi2}
    \chisq  =
   \left( \vv{M}_{\text{exp}} - \vv{M}_{\text{theo}} \right)^\mathrm{T}
               \covtot^{-1}
             \left( \vv{M}_{\text{exp}} - \vv{M}_{\text{theo}} \right),
\end{equation}
with the included measured moments $\vv{M}_{\mathrm{exp}}$, the corresponding theoretical prediction of these moments $\vv{M}_{\mathrm{theo}}$ and the total covariance matrix \covtot\ defined as the sum of the experimental (\covexp) and the theoretical (\covtheo) covariance matrix, respectively.

In the analysis of \cite{Aubert:2007BABARHadronicMoments} the theoretical prediction for the moments $\vv{M}_\text{HQE}$ are calculated perturbatively in a Heavy-Quark Expansion (HQE) in the kinetic-mass scheme up to $\mathcal{O}(1/m_b^3)$ with perturbative contributions \cite{Benson:2003kp,Gambino:2004qm,Aquila:2005hq} resulting in a dependence on six parameters: the running masses of the b- and c-quarks, $m_b(\mu)$ and $m_c(\mu)$, the parameters $\mupi$ and $\muG$ at $\mathcal{O}(1/m_b^2)$ in the HQE, and, at $\mathcal{O}(1/m_b^3)$, the parameters $\rhoD$ and $\rhoLS$.

New in this analysis is the inclusion of possible right-handed quark currents in the calculation of the theoretical prediction of the moments. The right-handed contributions are calculated and used here up to $\mathcal{O}(1/m_b^2)$ in the HQE and $\mathcal{O}(\alpha_s)$ in the perturbative correction. The aim of this fit is to give an upper bound for the relative contribution of a right-handed current compared with the standard-model left-handed current, which is parametrized by a prefactor $\cR$ for the new contributions to test. Thus the theoretical prediction of the moments depends on seven parameters to fit:
\[\vv{M}_\text{theo}=\vv{M}_\text{theo}(\cR, m_b, m_c, \mupi, \muG, \rhoD, \rhoLS).\]

\subsection{Determination of \boldmath\Vcb}

In the presence of a right-handed mixture the definition of the parameter \Vcb becomes ambiguous. Out of the three parameters \Vcb, \cL, \cR only two are independent, since \cL can be absorbed into \Vcb. To this end we choose to define
\[\Vcb\bar b_L\gamma_\mu c_L\quad\to\quad \Vcb\left(\bar b_L\gamma_\mu c_L+\cR\bar{b}_R\gamma_\mu c_R\right).\]

For the determination of \Vcb the fit uses a linearized form of the semileptonic rate \Gammasl expanded around \textit{a-priori} estimates of the HQE parameters \cite{Benson:2003kp}:

\begin{align}\label{eq:LinearVcb}
        \frac{\Vcb}{0.0417}  = & \sqrt{\frac{\brf_{clv}}{0.1032} \frac{1.55}{\tau_{\B}} }\bigl[1 + 0.30 \,(\alpha_s(\mb) - 0.22) \bigr] \\
               &\times
               \begin{aligned}[t]
                  \bigl[&1 - 0.66\, ( \mb - 4.60) + 0.39\, ( \mc - 1.15 ) \\
                  &+ 0.013\,( \mupi - 0.40) + 0.09\, ( \rhoD - 0.20) \\
                  &+ 0.05\, ( \muG - 0.35 ) - 0.01\, ( \rhoLS + 0.15 ) \\
                  &+ 0.341\, \cR\bigr].
               \end{aligned}\nonumber
\end{align}
Note the last term ($0.341\,\cR$), taking into account the possible contributions from a right-handed quark current. The \textit{a-priori} estimate of \cR is zero, i.\,e.\ the standard-model value. Due to the sizable factor and positive sign, a positive value of \cR increases \Vcb compared to the standard-model fit without \cR.

The total branching fraction $\brf(\semilepXc)$ in the fit is extrapolated from measured partial branching fractions $\brf_{\plmin}(\semilepXc)$, with $\plep \geq \plmin$. This is done by comparison with the HQE prediction of the relative decay fraction (r.h.s):
\begin{equation}
    \frac{\brf_{\plcut}(\semilepXc)}{\brf(\semilepXc)} = \frac{\int_{\plcut} \frac{\text{d}\Gammasl}{\text{d}\El} \text{d} \El}
                      {\int_{0} \frac{\text{d}\Gammasl}{\text{d}\El} \text{d} \El}.
\end{equation}
Thus the total branching fraction can be introduced as a free parameter in the fit. By adding the average $\B$-meson lifetime $\tau_{\B}$ (average between neutral and charged $B$-mesons, see also next paragraph) to the measured and predicted values, \Vcb can as well be introduced as a free parameter using \eqref{eq:LinearVcb}.

\begin{figure*}
   \hspace*{-5mm}\includegraphics{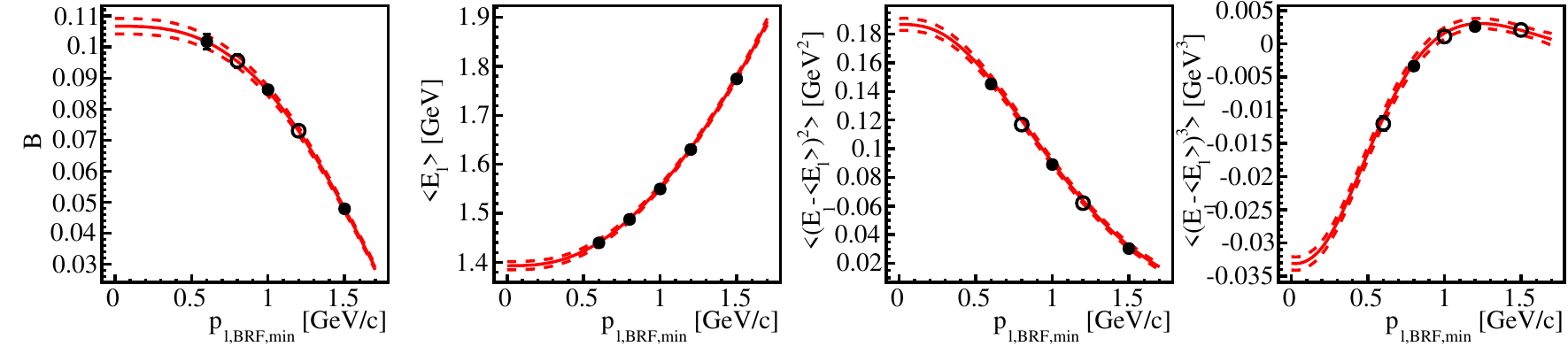}
   \caption{The measured lepton-moments (\textcolor{black}{$\bullet$}/\textcolor{black}{$\circ$}) compared with
            the result of the simultaneous fit (solid red line) as function of the
            minimal lepton momentum $\plmin$. The measurements included in the fit are marked by solid data
            points (\textcolor{black}{$\bullet$}). The dashed lines indicate the theoretical fit uncertainty obtained by the variation of the fit parameters in order to convert their theoretical uncertainty into an error of the moments.
           }
    \label{fig:FitLeptonMoments}
\end{figure*}

\begin{figure*}
   \hspace*{-5mm}\includegraphics{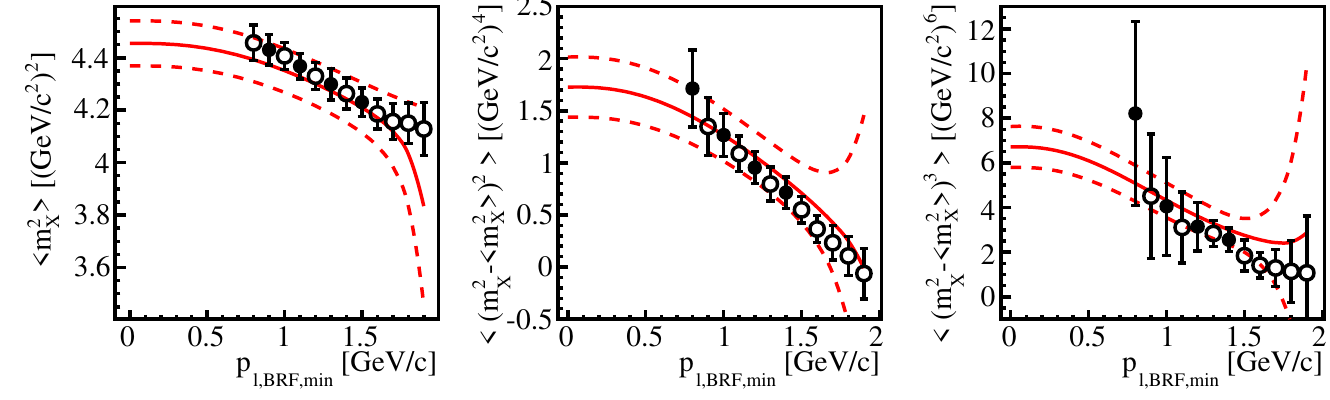}
   \caption{The measured hadronic-mass moments (\textcolor{black}{$\bullet$}/\textcolor{black}{$\circ$}) compared with the result of the simultaneous fit (solid red line) as function of the minimal lepton momentum $\plmin$. The measurements included in the fit are marked by solid data points (\textcolor{black}{$\bullet$}). The dashed lines indicate the theoretical fit uncertainty obtained by the variation of the fit parameters in order to convert their theoretical uncertainty into an error of the moments.
           }
    \label{fig:FitMassMoments}
\end{figure*}

\subsection{Experimental Input}
The combined fit is performed with a selection of the following 25 moment measurements by \babar\ which are characterized by correlations below $95\percent$ to ensure the invertibility of the covariance matrix:
\begin{itemize}
   \item
      Lepton energy moments measured by \babar\ \cite{Aubert:2004BABARLeptonMoments}.
      We use the partial branching fraction $\brf_{\plcut}$ at the minimal lepton momentum \linebreak\plgeq{0.6,1.0,1.5},
      the moments \elmom{} for \plgeq{0.6,0.8,1.0,1.2,1.5}, the central moments \elmomc{2} for \plgeq{0.6,1.0,1.5}
      and \elmomc{3} for \plgeq{0.8,1.2}.

   \item
      Hadronic mass moments measured by \babar\ \cite{Aubert:2007BABARHadronicMoments}. We use the
      moment \mxmom{2} for \linebreak\plgeq{0.9,1.1,1.3,1.5} and the central moments \mxmomc{2} and \mxmomc{3} both for \plgeq{0.8,1.0,1.2, 1.4}.
\end{itemize}
Furthermore we use the average $\B$ meson lifetime $\tau_{\B} = f_0 \tau_0 + (1 - f_0) \tau_{\pm} = (1.585 \pm 0.007) \ps$ with the lifetimes of neutral and charged $\B$ mesons $\tau_0$ and $\tau_{\pm}$ and the relative production rate, $f_0 = 0.491 \pm 0.007$, as quoted in \cite{Yao:2006pdbook}.

\subsection{Theoretical Uncertainties}
\label{TheoryErrors}
Theoretical uncertainties for the prediction of the moments $\vv{M}_{\mathrm{theo}}$ are estimated by variation of the parameters. The standard model parameters, that are all except \cR, are treated as in \cite{Aubert:2007BABARHadronicMoments}. The uncertainty in the non-perturbative part are estimated by varying the corresponding parameters \mupi and \muG by 20\percent and \rhoD and \rhoLS by 30\percent around their expected value. For the uncertainties of the perturbative corrections $\as=0.22$ is varied up and down by 0.1 for the hadronic mass moments and 0.04 for the lepton energy moments and the uncertainties of the perturbative correction of the quark masses $m_b$ and $m_c$ are estimated by varying them $20\,\text{MeV}/c^2$ up and down. An additional error of 1.4\percent is added to \Vcb from the fit for the uncertainty in the expansion of the semileptonic rate $\Gamma_\text{SL}$, which is not included in the fit, but quoted separately as theoretical uncertainty on \Vcb.

Additionally the influence of the right-handed contributions on the theoretical uncertainties in the predictions of the moments has to be included. Varying \cR in a similar fashion as the other parameters, around the a~priory estimate of zero showed only very little influence on the fit results. Due to the fact that the right-handed contributions are included up to $1/m_b^2$ in the non-perturbative and $\mathcal{O}(\alpha_s)$ in the perturbative corrections for all moments, the uncertainties in the prediction of the moments are not sizable and thus the variation of \cR has to be rather small. For the final results the variation of \cR has not been included, because of no influence on the significant digits.



\section{Results}
\label{sec:results}
\enlargethispage{5mm}
\begin{table*}
     \caption{Results of the full fit for \cR and the canonical set of parameters $\Vcb$, $\mb$, $\mc$, $\brf$, $\mupi$, $\muG$, $\rhoD$ and $\rhoLS$, separated by experimental and theoretical uncertainties. For $\Vcb$ we take into account an
additional error of $1.4\percent$ for the uncertainty in the expansion of the semileptonic
rate $\Gammasl$. Correlation coefficients for the parameters are listed below. The uncertainties $\Delta_\text{exp}$ and $\Delta_\text{theor}$ are the expected experimental and theory errors determined by Toy-MC studies (see \cite{Aubert:2009qda}) while $\Delta_\text{tot}$ is the total uncertainty
provided by the fit.
          }
     \begin{ruledtabular}

\begin{tabular}{lrrrrrrrrr}
 &$c^\prime_R$ &$\Vcb$ &$\mb$ &$\mc$ &$\brf$ &$\mupi$ &$\muG$ &$\rhoD$ &$\rhoLS$  \\
 & &$\times 10^{-3}$ &$[\gevcc]$ &$[\gevcc]$ &$[\%]$ &$[\gev^{2}]$ &$[\gev^{2}]$ &$[\gev^{3}]$ &$[\gev^{3}]$  \\
\hline 
Results &0.0517 &42.61 &4.588 &1.133 &10.674 &0.472 &0.303 &0.192 &-0.122  \\
$\Delta_\mathrm{exp}$ &0.1335 &2.00 &0.000 &0.000 &0.241 &0.032 &0.052 &0.015 &0.096  \\
$\Delta_\mathrm{theo}$ &0.4209 &5.55 &0.110 &0.149 &0.068 &0.094 &0.035 &0.056 &0.012  \\
$\Delta_\mathrm{\Gammasl}$ & &0.60 & & & & & & &  \\
$\Delta_\mathrm{tot}$ &$\asympm{+0.3356}{-0.4962}$ &$\asympm{+4.76}{-6.34}$ &$\asympm{+0.081}{-0.158}$ &$\asympm{+0.112}{-0.226}$ &$\asympm{+0.240}{-0.274}$ &$\asympm{+0.121}{-0.086}$ &$\asympm{+0.061}{-0.064}$ &$\asympm{+0.095}{-0.043}$ &$\asympm{+0.095}{-0.099}$  \\
\hline 
$c^\prime_R$ &1.00 &0.99 &0.78 &0.74 &0.67 &-0.79 &0.33 &-0.84 &0.21  \\
$\Vcb$ & &1.00 &0.75 &0.72 &0.72 &-0.77 &0.29 &-0.82 &0.22  \\
$\mb$ & & &1.00 &0.99 &0.64 &-0.72 &0.28 &-0.68 &0.17  \\
$\mc$ & & & &1.00 &0.65 &-0.70 &0.16 &-0.64 &0.20  \\
$\brf$ & & & & &1.00 &-0.47 &0.16 &-0.52 &0.14  \\
$\mupi$ & & & & & &1.00 &-0.21 &0.84 &-0.14  \\
$\muG$ & & & & & & &1.00 &-0.31 &0.03  \\
$\rhoD$ & & & & & & & &1.00 &-0.29  \\
$\rhoLS$ & & & & & & & & &1.00  \\
\end{tabular} 

     \end{ruledtabular}
     \label{tab:dfm_fitResults}
\end{table*}

\begin{table*}
     \caption{Results of the standard model fit with \cR fixed to zero.}
     \begin{ruledtabular}

\begin{tabular}{lrrrrrrrrr}
 &\phantom{0.0000} &$\Vcb$ &$\mb$ &$\mc$ &$\brf$ &$\mupi$ &$\muG$ &$\rhoD$ &$\rhoLS$  \\
 & &$\times 10^{-3}$ &$[\gevcc]$ &$[\gevcc]$ &$[\%]$ &$[\gev^{2}]$ &$[\gev^{2}]$ &$[\gev^{3}]$ &$[\gev^{3}]$  \\
\hline 
Results & &41.93 &4.578 &1.120 &10.654 &0.482 &0.300 &0.198 &-0.125  \\
$\Delta_\mathrm{exp}$ & &0.44 &0.058 &0.085 &0.175 &0.023 &0.040 &0.015 &0.081  \\
$\Delta_\mathrm{theo}$ & &0.38 &0.045 &0.067 &0.061 &0.055 &0.043 &0.027 &0.049  \\
$\Delta_\mathrm{\Gammasl}$ & &0.59 & & & & & & &  \\
$\Delta_\mathrm{tot}$ & &$\asympm{+0.83}{-0.83}$ &$\asympm{+0.074}{-0.070}$ &$\asympm{+0.106}{-0.107}$ &$\asympm{+0.185}{-0.185}$ &$\asympm{+0.058}{-0.060}$ &$\asympm{+0.059}{-0.059}$ &$\asympm{+0.031}{-0.031}$ &$\asympm{+0.094}{-0.094}$  \\
\hline 
$\Vcb$ & &1.00 &-0.35 &-0.21 &0.67 &0.29 &-0.39 &0.35 &0.06  \\
$\mb$ & & &1.00 &0.98 &0.25 &-0.26 &0.06 &-0.08 &0.01  \\
$\mc$ & & & &1.00 &0.29 &-0.29 &-0.11 &-0.07 &0.07  \\
$\brf$ & & & & &1.00 &0.14 &-0.08 &0.10 &0.00  \\
$\mupi$ & & & & & &1.00 &0.08 &0.52 &0.04  \\
$\muG$ & & & & & & &1.00 &-0.07 &-0.04  \\
$\rhoD$ & & & & & & & &1.00 &-0.20  \\
$\rhoLS$ & & & & & & & & &1.00  \\
\end{tabular} 

     \end{ruledtabular}
     \label{tab:sm_fitResults}
\end{table*}
\begin{figure*}
   \hspace*{-5mm}\includegraphics{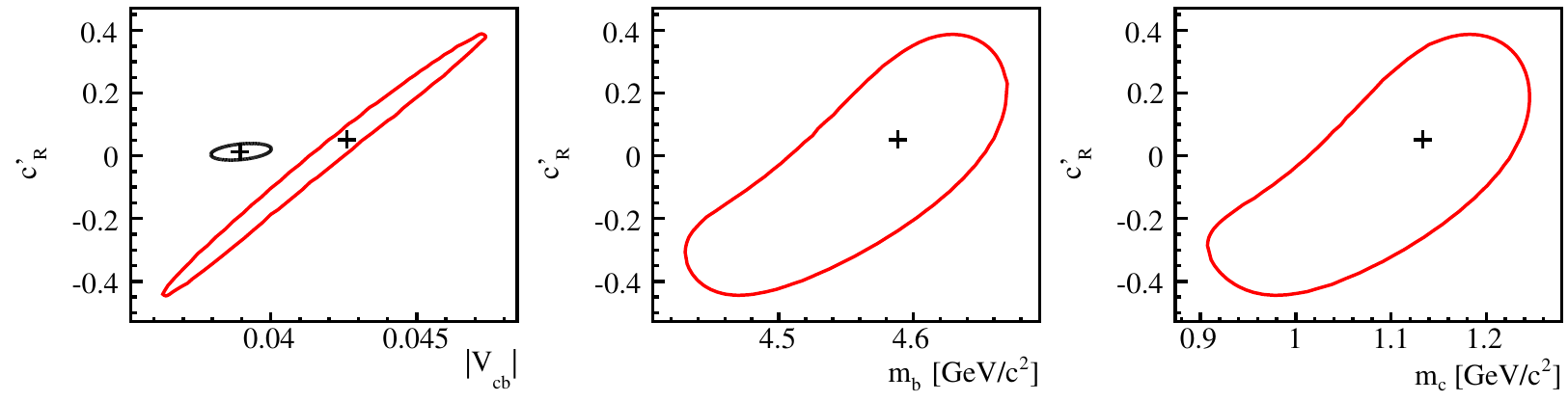}
   \caption{The $\Delta\chisq = 1$ contours in the $(\cR,\Vcb)$, $(\cR,m_b)$
            and $(\cR,m_c)$ plane for the results obtained in the fit.
            The small black contour in the $(\cR,\Vcb)$ plot shows the result from exclusive decays ($\cR = 0.01\pm0.03$) as computed in section \ref{sec:exclusive}.
           }
    \label{fig:FitContours}
\end{figure*}

Table \ref{tab:dfm_fitResults} shows the fit results and table \ref{tab:sm_fitResults} the corresponding standard-model fit results, which were obtained by performing the fit with \cR fixed to zero. Figs. \ref{fig:FitLeptonMoments} and \ref{fig:FitMassMoments} show a comparison of the fit results with the measured moments for the lepton-moments and the hadronic-mass moments, respectively.
The uncertainties $\Delta_\text{exp}$ and $\Delta_\text{theor}$ are the expected experimental and theory errors determined by Toy-MC studies (see \cite{Aubert:2009qda}) while $\Delta_\text{tot}$ is the total uncertainty
provided by the fit.

The estimate for $\cR=0.05\asympm{+0.33}{-0.50}$ is consistent with the standard-model prediction of zero, but the uncertainty reveals an unexpected low sensitivity of the semileptonic fit to possible right-handed contributions. We state the upper relative admixture limit of $|\cR|=0.9$
at 95\,\% confidence level.

The extracted value of $\Vcb=(43\asympm{+5}{-6})\cdot 10^{-3}$ is consistent with the value from the standard-model fit, but its uncertainty is quite different. In our fit, this is due to the influence of a sizable \cR uncertainty on the determination of \Vcb in \eqref{eq:LinearVcb}, which becomes evident in the contour plot of the $(\cR,\Vcb)$ plane, showing a shallow and steep covariance ellipse.

To compare the quality of the fits the P-value ($\text{prob}(\chi^2, n_\text{dof})$) suits best, because the fits differ by their number of degrees of freedom and thus the $\chi^2$ value alone is not sufficient. For the fit with \cR we find
$\chi^2=7.299$
with 17 degrees of freedom and thus
$\text{prob}(7.299, 17)=0.979$
and for the standard-model fit
$\text{prob}(7.312, 18)=0.987$%
, which shows neither improvement nor worsening.

The uncertainty of the result for \cR is dominated by the theory error $\Delta_\mathrm{theo}{=}0.42$ in table \ref{tab:dfm_fitResults} compared to $\Delta_\mathrm{exp}{=}0.13$. As a consequence including additional experimental data, e.\,g. from Belle, will not improve the limit for a possible right-handed contribution at this point. We investigated the theory error by changing the variation of the parameters as described in section \ref{TheoryErrors}. It turned out that the theory error cannot be pinned down to the uncertainty of a specific parameter. Furthermore, decreasing all theory errors had only little effect on the theory error of \cR. We come to the conclusion that the shape of the considered spectra and hence their moments are too similar for a left and right-handed $b\to c$ current, ending up in a low sensitivity of \cR and a weak constraint therein. Thus, improving the theoretical description, either for the standard-model part or the right-handed contributions, e.\,g.\ including the $1/m_b^3$ and BLM corrections of the right-handed current, will not reduce the uncertainty of \cR significantly in the performed moment analysis.

\section{Right-Handed Admixture from Exclusive Decays}
\label{sec:exclusive}
It is interesting to note that the value of \Vcb extracted in this way $\Vcb=(43\asympm{+5}{-6})\cdot 10^{-3}$ is in agreement with \Vcb from exclusive decays, from which \linebreak $\Vcb=(38.6{\pm}1.3)\cdot10^{-3}$ (from $\semilepDstar$) \cite{Amsler:2008zz} is obtained, while the value from the standard-model fit $\Vcb=(41.9{\pm}0.8)\cdot10^{-3}$ is not. This is due to the low sensitivity of \cR and thus the large uncertainty of the extracted value.
In addition, also the exclusive decays allow us to constrain a possible right-handed admixture \cite{Voloshin:1997zi, Mannel:1998ir}.

The most straightforward way of obtaining this information is to study the exclusive differential rates  at the
point of maximal momentum transfer to the leptons, corresponding to equal four-velocities of the initial and
final hadron. We consider the decays $\semilepD$ and $\semilepDstar$. The corresponding rates in the standard model are usually parametrized in terms of two form factors;
the relevant expressions close to the point of maximal momentum transfer read
\begin{align}
\frac{\mathrm{d} \Gamma^{B \to D}}{\mathrm{d} w} &= \Gamma_0 16 r^3 (r + 1)^2 (w^2 - 1)^{3/2} (\Vcb  {\cal G}(w))^2,\nonumber \\
\frac{\mathrm{d} \Gamma^{B \to D^*} }{\mathrm{d} w} &= \Gamma_0 192 r_*^3 (r_*-1)^2 (w^2 - 1)^{1/2} (\Vcb {\cal F}(w))^2
\end{align}
where $w {=} v{\cdot} v'$ is the scalar product of the hadronic velocities, $r{=}m_D/m_B$, $r_* {=} m_{D^*} / m_B$,
and $\Gamma_0 {=} G_F^2 m_B^5  / (192 \pi^3)$.

The information which is extracted by the experiments in the context of the $\Vcb^2$ determination is
\begin{align} \label{ExtData}
\lim_{w \to 1} \frac{\mathrm{d} \Gamma^{B \to D}}{\mathrm{d} w}  \frac{1}{ \Gamma_0 16 r^3 (r + 1)^2 (w^2 - 1)^{3/2} } ,\nonumber  \\
\lim_{w \to 1} \frac{\mathrm{d} \Gamma^{B \to D^*}}{\mathrm{d} w} \frac{1}{  \Gamma_0 192 r_*^3 (r_*-1)^2 (w^2 - 1)^{1/2} }
\end{align}
which in the standard model is the product of the form factors at $w = 1$ and \Vcb.  Combining
this with a theoretical prediction of the form factors at $w = 1$ one extracts \Vcb.

At the non-recoil point $w = 1$ the $B \to D$ transition is completely dominated by the vector current,
while the $B \to D^*$ decay is proportional to the axial vector current. Thus, including a right-handed
admixture,  the information extracted from (\ref{ExtData}) is $|\cL + \cRn| \Vcb {\cal G} (1)$ for the
case of the  $B \to D$ transition  and $|\cRn - \cL| \Vcb {\cal F} (1)$ for $B \to D^*$.  The current
experimental data yield \cite{Barberio:2008fa}:
\begin{align}
|\cL + \cRn| \Vcb {\cal G} (1) &= ( 42.4 \pm 1.56 ) \times 10^{-3}  \label{G1Vcb}\\
|\cRn - \cL| \Vcb {\cal F} (1)  &= (35.41 \pm 0.52 )  \times 10^{-3}\label{F1Vcb}
\end{align}

Using the lattice data (which are also used to extract \Vcb) \cite{Okamoto:2004xg, Gamiz:2008iv, Bernard:2008dn}
\begin{align}
   {\cal G}(1) &= 1.074 \pm 0.024  \label{G1}\\
   {\cal F}(1) &= 0.921 \pm 0.025  \label{F1}
\end{align}
we can extract the ratio $\cR = \cRn / \cL$ to be
\begin{equation}\label{ExclDet}
    \cR = 0.01\pm0.03
\end{equation}
with the assumption of no sizable correlations between the
      experimental measurements of the right-hand sides of Eqns. \eqref{G1Vcb} and
      \eqref{F1Vcb} as well as between the form factor values given in \eqref{G1} and
      \eqref{F1}.
The value for \Vcb extracted from Eqns. \eqref{G1Vcb} and \eqref{F1Vcb} is found to
      be $\Vcb\excl=(39.0\asympm{+1.1}{-1.0})\cdot 10^{-3}$ which has
      to be compared to $\Vcb\excl=(38.8 \pm 1.0) \cdot 10^{-3}$
      when setting $\cR=0$ in Eqns. \eqref{G1Vcb} and \eqref{F1Vcb}.

The result of \cR is compatible with zero and, in fact, more restrictive than the determination from
inclusive decays. Obviously the exclusive decay gives access to data separated by the handedness of the
$b\to c$ current in contrast to the inclusive decay, leading to a better limit on possible right-handed contributions.

In turn, we can use the result for \cR to determine $\Vcb\incl$ and compare with $\Vcb\excl$. This can be done by
imposing a Gaussian constraint of $\cR{=}0.01{\pm}0.03$ in the fit with a possible right-handed current.
The result $\Vcb\incl{=}(42.0{\pm}0.9){\cdot}10^{-3}$ compared to $\Vcb\excl{=}(39.0\asympm{+1.1}{-1.0}){\cdot}10^{-3}$ exhibits a tension by $3.0{\cdot}10^{-3}$ of the central values.
Determining \Vcb by \eqref{G1Vcb} and \eqref{F1Vcb} with \cR set to zero gives $\Vcb\excl(\cR{=}0){=}(38.8{\pm}1.0){\cdot}10^{-3}$ and allows us to
examine the differences in the tensions between inclusive and exclusive decays with and without a right-handed current, by comparing this value to the standard model fit value $\Vcb\incl(\cR{=}0){=}(41.9{\pm}0.8){\cdot}10^{-3}$ (see Table \ref{tab:sm_fitResults}) yielding a tension of about $3.1{\cdot}10^{-3}$ of the central values.
As a consequence, the difference in the central values between
      \Vcb exclusive and inclusive is slightly reduced, and more
      importantly, the uncertainty on \Vcb exclusive is considerably
      larger when allowing for a right-handed admixture resulting in a
      smaller significance of the observed effect. In our analysis,
      which is using only the inclusive \babar\ data, the difference between
      exclusive and inclusive is reduced from a $2.4\,\sigma$ to a
      $2.1\,\sigma$ effect.

\vspace{1cm}

\section{Summary}

\label{sec:summary}
We have performed a full-fledged fit to moments of the lepton-energy and hadronic-mass distribution of semileptonic $\bar{B} \to X_{c} \ell^{-} \bar{\nu}$ decays, including a possible right-handed admixture to the $b\to c$ current. We have considered the non-standard contributions up to $1/m_b^2$ in the non-perturbative and $\mathcal{O}(\alpha_s)$ in the perturbative corrections. The corresponding fit in the framework of the standard-model yields the most precise determination of \Vcb, due to the elaborated theoretical description and the precise measurements of the B factories \cite{Amsler:2008zz}. Our fit, including a right-handed admixture,  is in agreement with the standard model assumption of zero for a right-handed contribution. Unfortunately, the result $\cR=0.05\asympm{+0.33}{-0.50}$ reveals a low sensitivity of the fit to a right-handed contribution, compelling us to state the upper relative admixture limit of $|\cR|=0.9$
at 95\,\% confidence level. The moments of the spectra used in the fit are too similar for right- and left-handed contributions, resulting in the quoted low sensitivity and weak bound of \cR.

%

Exclusive decays are competitive in the determination of \Vcb, given the precise values for the form factors at the non-recoil point obtained from lattice QCD calculations.
The same precise values from lattice QCD calculations can be used to obtain a constraint on $\cR$, which is considerably stronger than the one obtained from inclusive decays:
$\cR {=} 0.01{\pm}0.03$. Using this result to determine $\Vcb\incl$ we can compare the tension between $\Vcb\incl$ and $\Vcb\excl$ without a right-handed current and with a right-handed current contribution as from exclusive decays. A right-handed current reduces the tension by about 3\,\% and its significance from a $2.4\,\sigma$ to a $2.1\,\sigma$ effect.
\vspace{-7mm}
%
%
%
$\quad$
\subsection*{Acknowledgements}
We acknowledge helpful discussions with I. Bigi, N. Uraltsev and
P. Giordano.
This work was partially supported by
the German Research Foundation (DFG) under contract No.
MA1187/10-1, and by the German Minister of Research (BMBF), contract No. 05HT6PSA.
\bibliographystyle{apsrev4-1}
\bibliography{DFMFit}

\end{document}